\documentclass[twocolumn,english,superscriptaddress,citeautoscript]{revtex4-2}
\usepackage[T1]{fontenc}
\usepackage[latin9]{inputenc}
\usepackage{color}
\usepackage{amsmath}
\usepackage{amssymb}
\usepackage{graphicx}
\usepackage{wasysym}
\usepackage{esint}

\makeatletter

\providecommand{\tabularnewline}{\\}

\usepackage{babel}

\usepackage{babel}

\makeatother

\usepackage{babel}
\begin{document}
\title{Magnetocaloric effect in $\mathrm{Cu}_{3}$-type compounds using the
Heisenberg antiferromagnetic model in a triangular ring}
\author{G. A. Antonio }
\affiliation{Department of Physics, Institute of Natural Science, Federal University
of Lavras, 37200-900 Lavras-MG, Brazil}
\author{J. Torrico}
\affiliation{Departamento de Física, Instituto de Ciências Exatas (ICEx), Universidade
Federal de Alfenas (UNIFAL), 37133-840 Alfenas, Minas Gerais, Brazil}
\author{A. S. da Mata}
\affiliation{Department of Physics, Institute of Natural Science, Federal University
of Lavras, 37200-900 Lavras-MG, Brazil}
\author{S. M. de Souza}
\affiliation{Department of Physics, Institute of Natural Science, Federal University
of Lavras, 37200-900 Lavras-MG, Brazil}
\author{Onofre Rojas}
\affiliation{Department of Physics, Institute of Natural Science, Federal University
of Lavras, 37200-900 Lavras-MG, Brazil}
\begin{abstract}
In this work we present a theoretical investigation into an antiferromagnetically
coupled spin system, specifically ${\rm Cu}_{3}-X$ ($\mathrm{X=As,Sb}$),
which exhibits an isosceles triangular configuration or slightly distorted
equilateral triangular configuration, as previously identified in
reference {[}Phys. Rev. Lett. \textbf{96}, 107202 (2006){]}. This
system can be effectively represented by the Heisenberg model on a
triangular structure, taking into account the exchange interaction,
the Dzyaloshinskii-Moriya interaction, g-factors and external magnetic
field, as delineated in the aforementioned reference. By using numerical
approach we explore both zero-temperature and finite-temperature behaviors
of a ${\rm Cu}_{3}$-like antiferromagnetically coupled spin system.
At zero temperature, the system displays a 1/3 quasi-plateau magnetization,
when the magnetic field is varied. Moreover, we place particular emphasis
on magnetic properties including magnetization, magnetic susceptibility,
entropy, and specific heat at finite temperatures. Furthermore, we
investigate the magnetocaloric effect as a function of an externally
imposed magnetic field, oriented both parallel and perpendicular to
the plane of the triangular structure. Interestingly, these configurations
demonstrate remarkably similar behavior for both orientations of the
magnetic field. Our investigation also includes an analysis of the
adiabatic curve, the Grüneisen parameter, and the variation in entropy
when applied or removed the magnetic field. The magnetocaloric effect
is found to be more prominent in low the temperature region, typically
at $T\sim1$K, for both parallel and perpendicular magnetic fields
at $\sim4.5$T and $\sim5$T, respectively.
\end{abstract}
\maketitle

\section{Introduction}

The study of spin systems with antiferromagnetic coupling has drawn
significant attention in the field of condensed matter physics. These
systems exhibit interesting features that arise from the interplay
of various factors that can be investigated through their magnetic
properties. Moreover, understanding the characteristics of these systems
helps to clarify their potential applications in areas such as magnetocaloric
materials and spintronics

The Magnetocaloric Effect (MCE) is a phenomenon that has been studied
extensively due to its potential applications in magnetic refrigeration
and cooling technologies. Initially observed in the late 19th century,
it refers to the change in temperature that occurs when a magnetic
material is subjected to a varying magnetic field, a phenomenon resulting
from the intrinsic magnetic properties of the material \citep{Tegus,Gruner}.
The reversibility of this effect has been confirmed in later studies,
sparking significant interest \citep{Gruner}. The MCE has been observed
in a variety of materials, such as rare earth alloys, magnetic oxides,
and transition metals, with notable instances of a giant magnetocaloric
effect (GMCE) driven by structural transitions \citep{Gschneidner,J-Liu}.
In 1951, Darby and colleagues made a pioneering step in the field
by designing a two-stage magnetocaloric regenerator using materials
with different Curie points, achieving temperature down to the final
values as low as 3mK at an induction of 0.42T \citep{darby}.

The concept of magnetic refrigeration at room temperature was introduced
almost a century after the discovery of MCE. In 1976, Brown developed
an efficient refrigeration system using gadolinium, marking a significant
advancement \citep{Brown}. Following this, in the late 90s, Gschneidner
discovered GMCE at room temperature in gadolinium-germanium-silicon
alloys (Ga-Ge-Si) \citep{Pecharsky}. Around the same time, Zimm proposed
a prototype showcasing the feasibility of magnetic refrigeration near
room temperature \citep{zimm}. These developments led to substantial
experimental and theoretical research on bulk $\mathrm{(Mn,Fe)_{2}(P,Si)}$-based
GMCE materials \citep{Pecharsky,Pecharsky01,ba-becerra,ba-zheng,F-zhang}.

Nanoscale materials with GMCE have gained attention due to their high
surface-to-volume ratio, enhanced interactions, and rapid thermal
response. These characteristics make them valuable for temperature
control applications. Examples of such applications include a room-temperature
thermal diode \citep{klinar}, a self-pumping magnetic cooling device
using Mn-Zn ferrite nanoparticles that achieves efficient energy conservation
without external energy input \citep{chaudhary}, a magnetic cooling
device based on a ferrofluid thermomagnetic that can effectively transfer
heat over large distances \citep{Pattanaik}, control of ferrofluid
droplets in microfluidics \citep{Sen}, and a magnetostructural phase
transition in Ni-Mn-Ga films showing a strong MCE at low magnetic
fields \citep{Qian}. Other applications involve gadolinium thick
films for energy conversion mechanisms \citep{ba-becerra,ba-zheng}
and biomedical applications like magnetic hyperthermia \citep{liu-zhang}
and efficient drug delivery via nanocarriers \citep{li-qu}.

Furthermore, the study of magnetic materials has attracted significant
attention due to their wide range of potential technological applications
in fields such as spintronics, nanoscale engineering, and biomedicine.
This has prompted investigations on $S=1/2$ antiferromagnetic triangular
spin rings, which might be ideal for observing peculiar quantum magnetization
due to two doublets. Compounds investigated include spin-frustrated
$(\mathrm{VO})_{3}^{6+}$-triangle-sandwiching octadecatungstates
as molecular magnets, displaying unusual magnetization jumps due to
predicted half-step or 1/3-plateau magnetization \citep{yamase}.
Experiments on a $\mathrm{Cu}_{3}$ nanomagnet revealed half-step
magnetization, hysteresis loops, and an asymmetric magnetization between
negative and positive field in fast sweeping external field, which
can be ascribed to an adiabatic change of mgnetization\citep{choi06}.
Whereas in \citep{trif} was investigated the spin-electric coupling.
The $S=1/2$ spin triangle clusters were also investigated, revealing
that the magnetization behavior and spin configurations are significantly
affected by the diamagnetic heteroatom ($X={\rm As}$ and ${\rm Sb}$)
\citep{choi08}. These clusters show potential for implementing spin-based
quantum gates \citep{choi12}. Bouammali et al. \citep{Bouammali}
explored the antisymmetric exchange in a tri-copper(II) complex, highlighting
its origins, theoretical implications, and potential for more advanced
electronic structure calculations. A spin-frustrated trinuclear copper
complex based on triaminoguanidine demonstrates strong antiferromagnetic
interactions with negligible antisymmetric exchange \citep{Spielberg}.
Several other studies have also examined triangular copper structures
\citep{belinsky,Robert,boudalis,stowe,kortz}.

On the other hand, theoretical investigations to explore various properties
of nanomagnets or magnetic molecular clusters, beyond experimental
results, are highly significant. For instance, Kowalewska and Sza\l owski
conducted a theoretical study of the magnetocaloric properties of
$V6$, a polyoxovanadate molecular magnet. Their research, using numerical
diagonalization and field ensemble formalism, uncovered highly tunable
magnetocaloric effects \citep{Szalowski}. Kar\v{I}ová et al. studied
the magnetization in antiferromagnetic spin-1/2 XXZ Heisenberg clusters,
demonstrating additional magnetization plateaux due to quantum interaction
and an enhanced magnetocaloric effect near magnetization shifts \citep{karlova}.
Reference \citep{torrico-20} employed exact diagonalization to examine
the spin-1/2 Hamiltonian for coupled isosceles Heisenberg triangles,
yielding a zero-temperature quantum phase transition diagram and a
magnetization profile. They also analyzed the thermodynamic behavior
and MCE. Another theoretical study was conducted on a $\mathrm{Cu}_{5}$
pentameric molecule using a spin-1/2 Heisenberg model, which explored
the thermodynamic properties, phase diagram, magnetization, and magnetocaloric
effects \citep{torrico-22}. Theoretical study of the MCE in paramagnetic
$\mathrm{PrNi}_{2}$ revealed an unexpected inverse effect due to
an anomalous increase in magnetic entropy at low temperatures \citep{Ranke}.
Several other theoretical investigations can be found in references
therein \citep{Szalowski,karlova,torrico-20,torrico-22,Ranke}.

In this context, a system of interest is ${\rm Cu}_{3}-X$ ($\mathrm{X=As,Sb}$),
which adopts an isosceles triangular or slightly distorted equilateral
triangular configuration. Previously, Choi et al. \citep{choi06,choi08,choi12}
have established that the behavior of this system can be effectively
described by the Heisenberg model on a triangular structure, incorporating
elements such as exchange interaction, Dzyaloshinskii-Moriya interaction,
g-factors, and external magnetic fields. Exploring the magnetic properties
and thermodynamic behavior of this ${\rm Cu}_{3}$-like spin system
is important as it helps us understand its fundamental characteristics
and identify potential advantages for its applications.

The article is organized as follows: in sec. 2 we present the model
and analyze some fundamental properties. In sec. 3 we explore main
thermodynamic properties. Whereas in sec.4 we discuss the magnetocaloric
effect. Finally in sec. 5 we devote our conclusions.

\section{Model}

In this work, we aim to explore the thermodynamics and magnetic properties
of a triangular cluster ${\rm Na}_{9}[{\rm Cu}_{3}{\rm Na}_{3}({\rm H}_{2}{\rm O})_{9}(\alpha-X{\rm W}_{9}{\rm O}_{33})_{2}]$
(where $X={\rm As}$ and ${\rm Sb}$) hereinafter referred to as the
\{${\rm Cu}_{3}-X$\} system\citep{choi06}. The compound under consideration
contains three copper atoms, each of which loses two electrons to
form a Cu(II) or $\mathrm{Cu}^{+2}$ ion. The electron loss in Cu(II)
ions occurs from both the 4s and one of the 3d orbitals, resulting
in a single unpaired electron and a net magnetic moment with a spin
of $S=1/2$; this behavior can be adequately described by the Heisenberg
model within the framework of an isosceles triangular spin ring \citep{choi06,choi08,choi12}.
Consequently, we adopt the Hamiltonian, as presented in previous work
\citep{choi06,choi08,choi12}, which characterizes ${\rm Cu}_{3}$-like
compounds, as follows 
\begin{alignat}{1}
\mathbf{H}= & \sum_{j=1}^{3}\sum_{\alpha=x,y,z}J_{j,j+1}^{\alpha}S_{j}^{\alpha}S_{j+1}^{\alpha}\nonumber \\
 & +\sum_{j=1}^{3}\Bigl[{\bf D}_{j,j+1}\cdot\left({\bf S}_{j}\times{\bf S}_{j+1}\right)+\mu_{B}{\bf S}_{j}\cdot\mathbf{g}_{j}\cdot{\bf B}_{j}\Bigr],\label{eq:H1}
\end{alignat}
where $S_{j}^{\alpha}$ denotes the spin-1/2 components of localized
${\rm Cu}_{3}$-like spin with $\alpha=\{x,y,z\}$, and $J_{j,j+1}^{\alpha}$
(simplified as $J_{j}^{\alpha}$) represents the exchange interaction
parameters between site $j$ and $j+1$ for each component (for schematic
view see reference \citep{choi06,choi08,choi12}). The second term
refers to the Dzyaloshinskii-Moriya interaction vector ${\bf D}_{j,j+1}$
denoted as ${\bf D}_{j,j+1}=(D_{j,j+1}^{x},D_{j,j+1}^{y},D_{j,j+1}^{z})$.
The site-dependent $g$-factors are defined as $\mathbf{g}_{j}=(g_{j}^{x},g_{j}^{y},g_{j}^{z})$,
while the last term corresponds to the magnetic field $\mathbf{B}$,
which we assume to be independent of the spin site on the triangle.
Here, $\mu_{B}$ denotes the Bohr magneton. The specific parameters
were obtained using Electron Spin Resonance (ESR) data \citep{choi06,choi08,choi12},
and these parameters are reproduced in table \ref{tab:1} for both
compounds. It is worth mentioning that only ${\bf D}_{1,2}=(D,D,D)$
is isotropic, while ${\bf D}_{2,3}$ and ${\bf D}_{3,1}$ contribute
solely to the $z$-component, expressed as ${\bf D}_{2,3}={\bf D}_{3,1}=(0,0,D)$.
Other interactions, such as the crystal field effect and magneto-crystalline
anisotropy, were disregarded in this study because their contributions
are not deemed highly relevant, as supported by references\citep{choi06,kortz}.

For convenience, we express the Hamiltonian \eqref{eq:H1} in units
of kelvin (K). Hence, let us redefine $\mu_{B}$ as $\hat{\mu}_{B}=\frac{\mu_{B}}{k_{B}}=0.6717156644$
K/T, where $k_{B}$ denotes the Boltzmann constant. Therefore, the
magnetic field $\mathbf{B}$ is conveniently measured in tesla (T)
units. This can be equivalent to setting the Boltzmann constant as
$k_{B}=1$, implying that, for the sake of simplicity, all calculation
will be expressed in units of $k_{B}$.

\begin{table}[h]
\begin{tabular}{|r|r|c|c|}
\hline 
Magnetic parameters & notation & \{${\rm Cu}_{3}-{\rm As}$\} & \{${\rm Cu}_{3}-{\rm Sb}$\}\tabularnewline
\hline 
\hline 
$J_{1,2}^{x}=J_{1,2}^{y}$ & $J_{1}$ & $4.50$ K & $4.49$ K\tabularnewline
\hline 
$J_{1,2}^{z}$ & $J_{1}^{z}$ & $4.56$ K & $4.54$ K\tabularnewline
\hline 
$J_{2,3}^{x}=J_{2,3}^{y}=J_{3,1}^{x}=J_{3,1}^{y}$ & $J_{2}$ & $4.03$ K & $3.91$ K\tabularnewline
\hline 
$J_{2,3}^{z}=J_{3,1}^{z}$ & $J_{2}^{z}$ & $4.06$ K & $3.96$ K\tabularnewline
\hline 
$D_{1,2}^{z}=D_{2,3}^{z}=D_{3,1}^{z}$ & $D$ & $0.529$ K & $0.517$ K\tabularnewline
\hline 
$D_{1,2}^{x}=D_{1,2}^{y}$ & $D$ & $0.529$ K & $0.517$ K\tabularnewline
\hline 
$g_{1}^{x}=g_{1}^{y}$ & $g_{1}$ & $2.25$ & $2.24$\tabularnewline
\hline 
$g_{2}^{x}=g_{2}^{y}$ & $g_{2}$ & $2.10$ & $2.11$\tabularnewline
\hline 
$g_{3}^{x}=g_{3}^{y}$ & $g_{3}$ & $2.40$ & $2.40$\tabularnewline
\hline 
$g_{1}^{z}=g_{2}^{z}=g_{3}^{z}$ & $g_{z}$ & $2.06$ & $2.07$\tabularnewline
\hline 
\end{tabular}\caption{\label{tab:1}Magnetic parameters of the \{${\rm Cu}_{3}-X$\} compounds,
where $X$ denotes either As or Sb, as extracted from reference\citep{choi08}.}
\end{table}

\section{Thermodynamics quantities }

The eigenvalues of the above mentioned Hamiltonian \eqref{eq:H1},
can be obtained by direct numerical diagonalization. More details
about the energy spectra can be found in reference \citep{choi06,choi08,choi12},
so let us assume that the eigenvalues can be expressed as follows
\begin{equation}
\mathbf{U}\mathbf{H}\mathbf{U}^{-1}=\mathbf{E}={\rm diag}\left(\varepsilon_{1},\varepsilon_{2},\cdots,\varepsilon_{8}\right),
\end{equation}
where $\mathbf{U}$ is an $8\times8$ matrix that diagonalizes the
Hamiltonian \eqref{eq:H1}. It is important to note that this matrix,
which naturally depends on the Hamiltonian parameters, can only be
obtained numerically for a fixed magnetic field.

Thus, the partition function can symbolically be represented by:
\begin{equation}
\mathcal{Z}={\rm tr}\left({\rm e}^{-\mathbf{E}/T}\right)=\sum_{i=1}^{8}{\rm e}^{-\varepsilon_{i}/T}.\label{eq:Zp}
\end{equation}
Here, the eigenvalues $\varepsilon_{i}$ (in kelvin units) depend
on the Hamiltonian parameters provided in table \ref{tab:1}, as well
as the magnetic field $\mathbf{B}$ (in tesla), while $T$ represents
the temperature of the system (in kelvin). In theory, any physical
quantity can be derived from the partition function \eqref{eq:Zp}.
However, as the eigenvalues can only be obtained numerically, physical
quantities that require derivatives, such as magnetization and magnetic
susceptibility among others, must be calculated with caution. Numerical
derivatives may not always provide accurate results, hence it is advisable
to avoid them as much as possible. Therefore, we will combine numerical
and analytical calculations to safely obtain all physical quantities.

In this regard, the free energy can be denoted by the expression
\begin{equation}
f=-T\ln(\mathcal{Z}).
\end{equation}
It should be noted that the free energy is also represented in units
of $k_{B}$.

\subsection{Internal energy }

The first quantity we will discuss is the internal energy, as it directly
influences the magnitude of the magnetocaloric effect. As previously
stated, the eigenvalues of the Hamiltonian can be obtained using the
parameters listed in table \ref{tab:1} and a fixed magnetic field.
Formally, the average internal energy can be represented as:
\begin{equation}
U=\langle\mathbf{H}\rangle=\frac{1}{\mathcal{Z}}{\rm tr}\left\{ \mathbf{H}{\rm e}^{-\mathbf{H}/T}\right\} =\frac{1}{\mathcal{Z}}\sum_{i=1}^{8}\varepsilon_{i}{\rm e}^{-\varepsilon_{i}/T}.
\end{equation}
For the purposes of our discussion, we will focus on the ${\rm Cu}_{3}-{\rm As}$
compound. The ${\rm Cu}_{3}-{\rm Sb}$ compound exhibits analogous
characteristics because the parameters given in table \ref{tab:1}
are quite similar.

\begin{figure}
\includegraphics[scale=0.7]{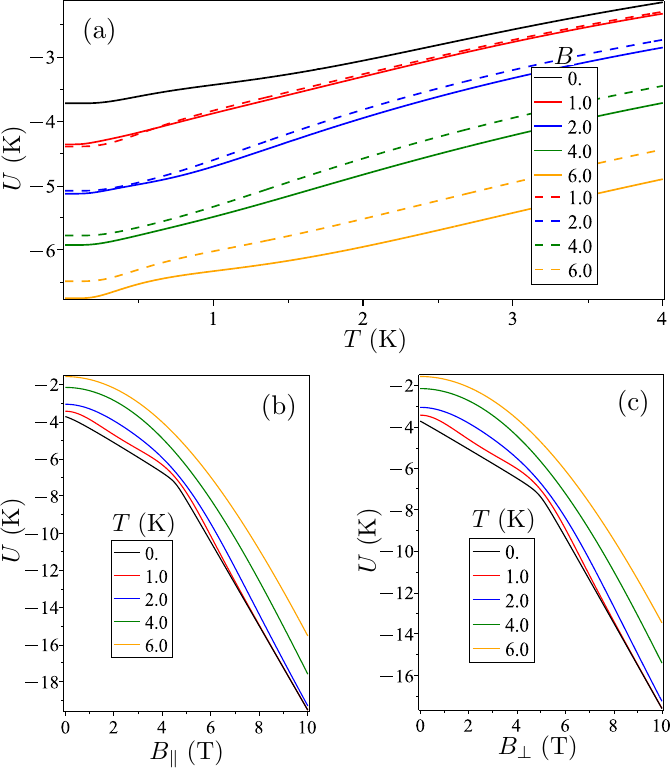}

\caption{\label{fig:U-TB}(a) Internal energy $U$ as a function of temperature,
solid lines corresponds to $B_{\perp}$, while dashed lines corresponds
to $B_{\parallel}$. (b-c) Internal energy $U$ as a function of perpendicular
and parallel external magnetic field, respectively. These plots are
specifically for the ${\rm Cu}_{3}-{\rm As}$ compound.}

\end{figure}

Figure \ref{fig:U-TB}a depicts the internal energy ($U$) as a function
of temperature, assuming a constant parallel external magnetic field
to the triangle plane (solid line) and perpendicular to it (dashed
line). The internal energy varies slightly between the parallel and
perpendicular magnetic fields. Although, as the magnetic field increases,
the discrepancy becomes more pronounced. In contrast, panels (b and
c) show the internal energy as a function of the parallel and perpendicular
external magnetic fields, respectively. These figures assume several
fixed temperatures, as specified inside the panel. At zero temperature,
we observe a significant change of internal energy at $B_{\parallel}\approx4.5$
T. Above this magnetic field, the system aligns entirely parallel
to the magnetic field, while for $B_{\parallel}\lesssim4.5$ T, the
configuration comprises two aligned spins and a third with opposite
alignment. We observe similar behavior when the external magnetic
field acts perpendicularly to the triangular plane, but the shift
occurs at a slightly higher magnetic field, $B_{\perp}\approx5$ T.
This similarity has previously been observed in energy spectra and
zero-temperature magnetization \citep{choi06,choi08,choi12}. As temperature
increases, this curvature smooths out. In the absence of an external
magnetic field, these spin moments are oriented randomly, leading
to a higher internal energy state. When an external magnetic field
is applied, the spins align with the field, reducing the internal
energy of the compound. This variation of energy manifests as a change
in the compound temperature, representing the core of the magnetocaloric
effect.

\subsection{Entropy}

Entropy calculation is relevant as it plays a crucial role in the
MCE. Essentially serving as the \textquotedbl driving force\textquotedbl{}
to understand both the direct and inverse MCE, which we will discuss
subsequently. As such, entropy is fundamental to understand the mechanism
of MCE and is pertinent in applications such as magnetic refrigeration.
Consequently, entropy can be derived from the internal energy with
the following relation
\begin{equation}
\mathcal{S}=\frac{\langle\mathbf{H}\rangle-f}{T}.
\end{equation}

\begin{figure}
\includegraphics[scale=0.6]{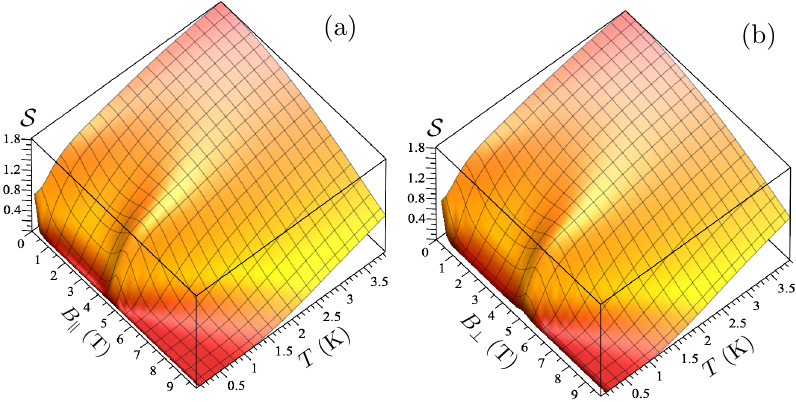}\caption{\label{fig:Entropy}(a) Entropy ${\cal S}$ in the plane of temperature
(in kelvin) and parallel external magnetic field (tesla). (b) Entropy
${\cal S}$ in the plane of temperature (in kelvin) and perpendicular
external magnetic field (tesla). These visualizations are based on
the ${\rm Cu}_{3}-{\rm As}$ compound.}
\end{figure}

In Fig. \ref{fig:Entropy}a the entropy is illustrated in the plane
of temperature (in kelvin) and parallel external magnetic field (tesla).
It is wort to note that $B_{\parallel}\approx4.5$T where the entropy
increases very fast in the low temperature region, this is because
the region dominated by two spins aligned and one spin aligned oppositely,
changes to a region with all aligned spins to magnetic field. Similarly,
in panel (b) is illustrated the entropy as in the plane of temperature
and perpendicular external magnetic field. Although the plot is quite
similar to panel (a) there are some slight difference like a strong
change occurs at little bit higher magnetic field $B_{\perp}\approx5$T.
It is worth to mention also in absence of external magnetic field
the system is two-fold degenerate, so the entropy leads to $\mathcal{S}\rightarrow\ln(2)$. 

\subsection{Specific heat}

Specific heat is of significant importance to the magnetocaloric effect
(MCE) as it fundamentally influences the amount of heat absorbed or
released during the application or removal of a magnetic field. It
quantifies the amount of heat required to change a compound temperature
by a certain amount. Therefore, we can use the following relation
to obtain the specific heat:
\begin{equation}
C=\frac{\langle\mathbf{H}^{2}\rangle-\langle\mathbf{H}\rangle^{2}}{T^{2}},
\end{equation}
where
\begin{equation}
\langle\mathbf{H}^{2}\rangle=\frac{1}{Z}\sum_{i=1}^{8}\varepsilon_{i}^{2}{\rm e}^{-\varepsilon_{i}/T}.
\end{equation}
It is worth to mention that the specific heat can depend analytically
of temperature, after eigenvalues is found numerically.

\begin{figure}
\includegraphics[scale=0.6]{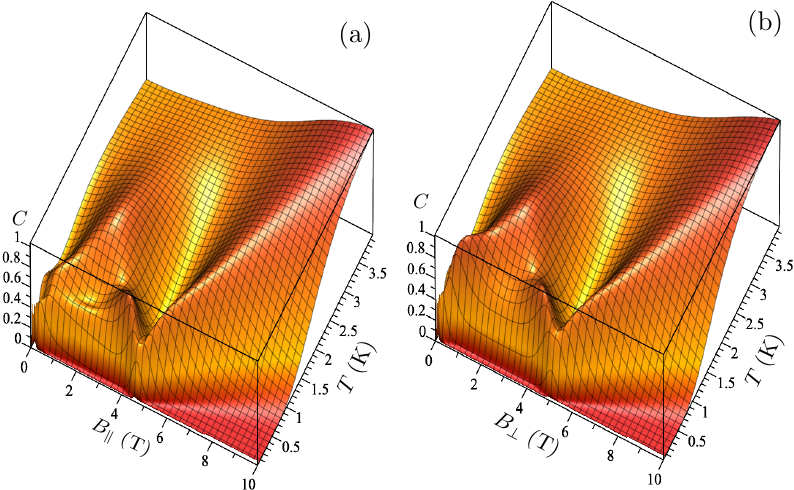}\caption{\label{fig:Specific-heat}(a) Specific heat $C$ as a function of
temperature and parallel external magnetic field. (b) Specific heat
$C$ as a function of temperature and perpendicular external magnetic
field. For the ${\rm Cu}_{3}-{\rm As}$ compound.}
\end{figure}

Figure \ref{fig:Specific-heat}a presents the specific heat as a function
of temperature and the parallel external magnetic field $B_{\parallel}$.
An anomalous behavior is noticeable at $B_{\parallel}\approx4.5$
T, which manifests as an unusual peak in the low-temperature region.
Additionally, two peculiar peaks appear at $B_{\parallel}\thicksim2$
T, whereas other regions with a fixed magnetic field exhibit only
one anomalous peak. Panel (b) illustrates an analogous plot, albeit
with the perpendicular magnetic field $B_{\perp}$. The specific heat
plots mainly resemble those in panel (a), with the exception of the
absence of a minimum at $B_{\perp}\thicksim2$ T. The low-temperature
anomaly occurs at $B_{\parallel}\approx5$ T. For a system with temperature
around $T\sim1$ K, the specific heat exhibits unusual behavior, absorbing
or releasing heat more efficiently for a given temperature change.
High specific heat is a beneficial property as it can enhance the
overall efficiency and effectiveness of the triangular system. From
the perspective of the MCE, this translates into more efficient magnetic
cooling or heating.

\subsection{Magnetization}

We will now discuss magnetization, which has a key role in the MCE.
The strength of the MCE is directly related to the change in magnetization
of the compound in response to variation in temperature and the applied
magnetic field. In our case, the magnetization can be derived without
taking numerical derivatives, through the following relation

\begin{equation}
\left\langle \left(\tfrac{\partial\mathbf{H}}{\partial B_{k}}\right)\right\rangle =\frac{1}{Z}{\rm tr}\left\{ \mathbf{H}_{B_{k}}{\rm e}^{-\mathbf{H}/T}\right\} =\frac{1}{Z}{\rm tr}\left\{ \tilde{\mathbf{H}}_{B_{k}}{\rm e}^{-\mathbf{E}/T}\right\} ,
\end{equation}
where $\tilde{\mathbf{H}}_{B_{k}}=\mathbf{U}\left(\tfrac{\partial\mathbf{H}}{\partial B_{k}}\right)\mathbf{U}^{-1}$,
and $k=\{\Vert,\bot\}$. This method is a typical procedure to avoid
numerical derivatives, as the Hamiltonian can be derived in relation
to $B_{k}$ analytically. Therefore, the magnetization becomes

\begin{equation}
M_{k}=-\frac{1}{\mathfrak{g}_{k}}\left\langle \left(\tfrac{\partial\mathbf{H}}{\partial B_{k}}\right)\right\rangle ,
\end{equation}
where $\mathfrak{g}_{k}$ is a constant normalization chosen for convenience,
defined as follows: $\mathfrak{g}_{k}=\frac{1}{3}\sum_{i=1}^{3}g_{i}^{k}$.
The values of $\mathfrak{g}_{\parallel}$ is $2.25$ for both compounds,
while the values of $\mathfrak{g}_{\perp}$ are $2.06$ and $2.07$
for As and Sb, respectively.
\begin{figure}
\includegraphics[scale=0.58]{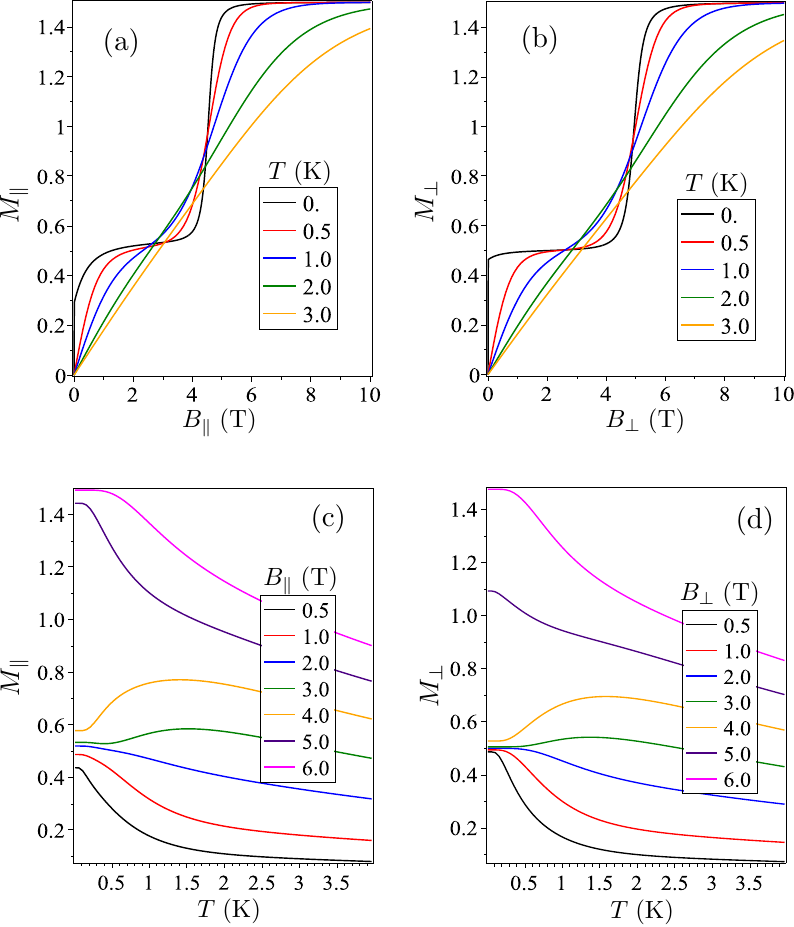}\caption{\label{fig:Mags}\textcolor{red}{{} }(a) Magnetization versus parallel
magnetic field at various temperatures. (b) Magnetization versus perpendicular
magnetic field at the same temperatures. (c) Temperature-dependent
magnetization for different parallel magnetic fields for several parallel
magnetic fields. (d) Analogously, temperature-dependent magnetization
for a set of perpendicular magnetic fields. The study considers the
${\rm Cu}_{3}-{\rm As}$ compound.}
\end{figure}

Figure \ref{fig:Mags}a presents the magnetization as a function of
the parallel magnetic field, $B_{\parallel}$, at various temperature
values, including zero-temperature. Note that a $1/3$ quasi-plateau
feature appears, fading as temperature increases to around $1$K.
Similarly, panel (b) reports magnetization as a function of a perpendicular
external magnetic field. Here, the $1/3$ quasi-plateau becomes more
noticeable, with effects largely mirroring those in the previous panel.
Conversely, panel (c) displays magnetization as a function of temperature,
considering multiple external magnetic fields parallel to the triangle
plane. In this panel, the quasi-plateau region converges to $M\sim0.5$,
with a saturated region at $M\rightarrow1.5$. A significant curvature
change occurs at around $1$ K. Lastly, panel (d) illustrates the
magnetization as a function of temperature for an external magnetic
field perpendicular to the triangular plane. The magnetization behavior
closely resembles that in panel (c), but with a more noticeable convergence
to the $1/3$ quasi-plateau in low-temperature regions. Here again,
the main curvature change happens approximately at $1$ K.

\subsection{Magnetic Susceptibility}

Magnetic susceptibility is another relevant quantity for studying
the MCE as it determines how easily a material can be magnetized or
demagnetized. To obtain this quantity, we can follow a procedure similar
to the previous one. Thus the magnetic susceptibility, can be given
by

\begin{equation}
\chi_{k}=\frac{1}{\mathfrak{g}_{k}^{2}T}\left\{ \left\langle \left(\tfrac{\partial\mathbf{H}}{\partial B_{k}}\right)^{2}\right\rangle -\left\langle \left(\tfrac{\partial\mathbf{H}}{\partial B_{k}}\right)\right\rangle ^{2}\right\} .
\end{equation}

\begin{figure}
\includegraphics[scale=0.6]{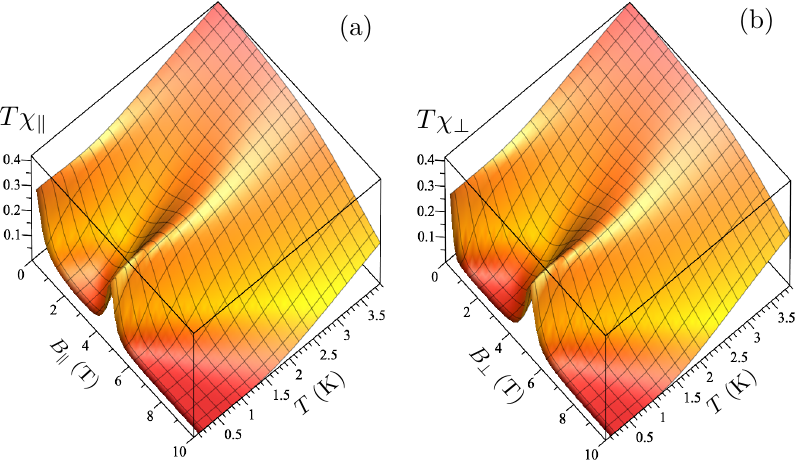}\caption{\label{fig:Magnetic-susceptibility}(a) Magnetic susceptibility times
temperature, $T\chi_{\parallel}$ in the plane of temperature and
parallel external magnetic field. (b) Magnetic susceptibility $T\chi_{\perp}$
in the plane of temperature and perpendicular external magnetic field.}
\end{figure}

Figure \ref{fig:Magnetic-susceptibility}a illustrates the magnetic
susceptibility times temperature ($T\chi_{\parallel}$) for ${\rm Cu}_{3}-{\rm As}$
compound, plotted against temperature (in kelvin) and parallel external
magnetic field (tesla). It is noteworthy that at $B_{\parallel}\approx4.5$
T, $T\chi_{\parallel}$ maintains a constant value of around $T\chi_{\parallel}\approx0.3$.
This means that the magnetic susceptibility inversely depends on the
temperature in the low-temperature region. An alike finite value is
observed for the null magnetic field. This is due to the shift from
regions dominated by two aligned spins and one oppositely aligned
spin, leading to complete alignment with the external magnetic field.
The ${\rm Cu}_{3}-X$ compound, with higher magnetic susceptibility,
can be magnetized or demagnetized more readily, resulting in greater
thermal energy transfer and a more significant temperature change,
roughly at $T\lesssim1$ K. Similarly, panel (b) depicts the product
of magnetic susceptibility and temperature, $T\chi_{\perp}$, in the
plane of temperature and perpendicular external magnetic field. Although
the behavior is quite similar to panel (a), there are slight differences,
such as the pronounced change occurring at a slightly higher magnetic
field, $B_{\perp}\approx5$ T. Therefore, magnetic susceptibility
affects the magnitude of the temperature change observed during the
MCE and could play a crucial part in enhancing the magnetic refrigeration
systems.

\section{Magnetocaloric Effect}

The Magnetocaloric Effect (MCE) refers to the thermal response of
a material to the change in an external magnetic field. It holds potential
for practical applications such as energy-efficient cooling technologies.
Accordingly, we will discuss aspects like the isentropic curve and
Grüneisen parameter.

\subsection{Isentropic curve}

In magnetic systems, isentropic curves or adiabatic temperature curves
provide a useful tool to visualize and understand the MCE. In the
context of MCE, an isentropic curve represents a process that occurs
at constant entropy in a magnetic field-temperature phase diagram. 

\begin{figure}
\includegraphics[scale=0.6]{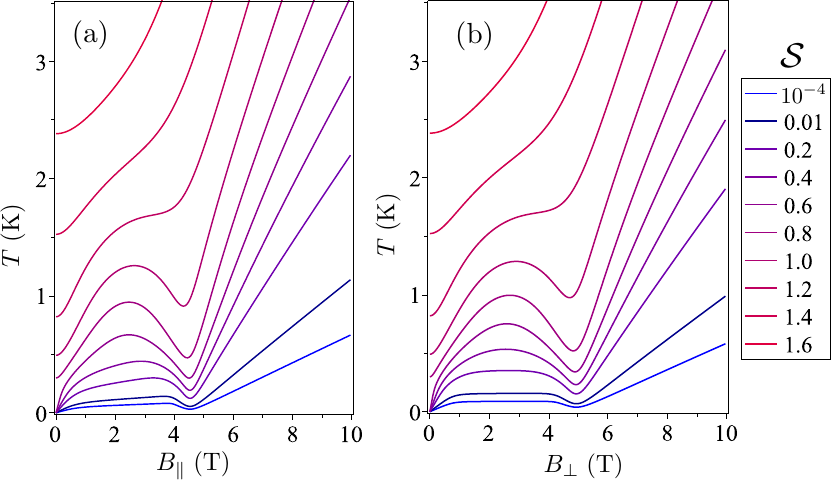}\caption{\label{fig:Isentp-As}(a) Plot of isentropic $\mathcal{S}$ (in units
of $k_{B})$ curve temperature $T\:({\rm K})$ against a parallel
magnetic field $B_{\parallel}\;({\rm T})$. (b) Plot of isentropic
curve temperature $T\:({\rm K})$ as a function of magnetic field
$B_{\perp}\;({\rm T})$. Both plots are for the ${\rm Cu}_{3}-{\rm As}$
compound.}
\end{figure}

In Figure \ref{fig:Isentp-As}a, the isentropic curve for \{${\rm Cu}_{3}-{\rm As}$\}
is illustrated for a parallel magnetic field. Between null magnetic
field and $B_{\parallel}\approx4.5$ T, the system exhibits the first
step of magnetization, with two spins aligned parallel to the magnetic
field while the third one is aligned antiparallel. Surely, for $B_{\parallel}\apprge4.5$
T, all spins become aligned with the magnetic field. The isentropic
curve shows high sensitivity at relatively low entropy, and as temperature
increases, the crossover region between these two states manifests
as a minimum. This minimum gradually disappears around $T\sim1$ K.
Notably, strong slopes of the isentropic curves occur around the minimum,
indicating a large MCE in this region. Similarly, in Figure \ref{fig:Isentp-As}b,
the isentropic curve for \{${\rm Cu}_{3}-{\rm As}$\} is depicted
for a perpendicular magnetic field. The behavior of the isentropic
curve is largely analogous to the previous case, with the only difference
being that the minimum occurs at approximately $B_{\perp}\approx5$
T. These curves provide insight into the temperature changes of a
system in response to the application or removal of an external magnetic
field. Additionally, the shape of the isentropic curve can provide
valuable information about possible zero-temperature magnetic phase
transitions in the \{${\rm Cu}_{3}-{\rm As}$\} compound.

\subsection{Grüneisen parameters}

The Grüneisen parameter plays a essential role in understanding the
MCE, which refers to the change in temperature of a material resulting
from variations in an applied magnetic field. This effect has significant
applications in magnetic cooling technologies. The Grüneisen parameter
quantifies the relationship between the change in temperature of the
compound and the magnetic field under constant entropy conditions.
Specifically, the Grüneisen parameter $\Gamma$ is defined as the
ratio of the temperature derivative of the magnetization per mole
to the molar specific heat,

\begin{equation}
\Gamma_{k}=-\frac{1}{C}\!\left(\frac{\partial M}{\partial T}\right)_{B_{k}}\!=-\frac{\left(\tfrac{\partial\mathcal{S}}{\partial B_{k}}\right)_{T}}{T\left(\tfrac{\partial\mathcal{S}}{\partial T}\right)_{B_{k}}}=\frac{1}{T}\!\left(\frac{\partial T}{\partial B_{k}}\right)_{\mathcal{S}}\!,
\end{equation}
To avoid numerical derivatives, an equivalent expression of the Grüneisen
parameter can be derived 

\begin{equation}
\Gamma_{k}=\frac{1}{\mathfrak{g}_{k}}\frac{\langle\mathbf{H}_{B_{k}}\mathbf{H}\rangle-\langle\mathbf{H}_{B_{k}}\rangle\langle\mathbf{H}\rangle}{\langle\mathbf{H}^{2}\rangle-\langle\mathbf{H}\rangle^{2}}.
\end{equation}
 Further research has explored these phenomena by analyzing the adiabatic
temperature change, which demonstrates a strong correlation with the
magnetic entropy change\citep{Zhu}. 

\begin{figure}
\includegraphics[scale=0.58]{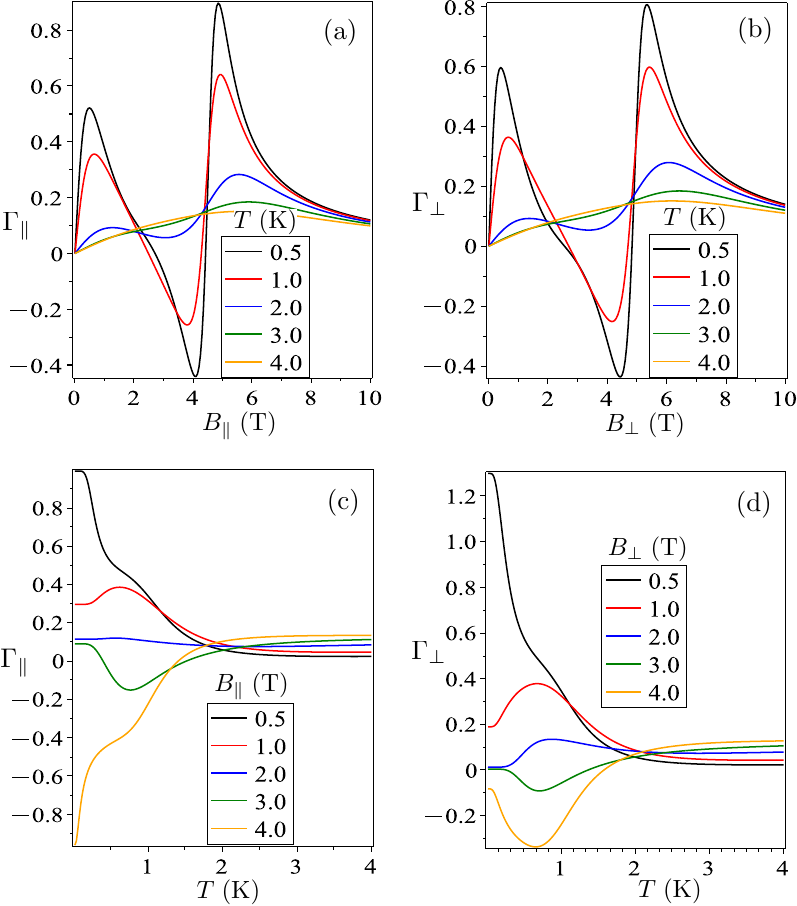}\caption{\label{fig:Gruneisn}(a) Grüneisen parameter $\Gamma_{\parallel}$
as a function of parallel magnetic field $B_{\parallel}$, for a range
of temperatures. (b) Grüneisen parameter $\Gamma_{\perp}$ as a function
of perpendicular magnetic field. (c) $\Gamma_{\parallel}$ as a function
of temperature for a set of parallel magnetic field. (d) $\Gamma_{\perp}$
as a function of temperature for a number of perpendicular magnetic
field. The magnetic field is in units of tesla, while temperature
is measured in kelvin. For the ${\rm Cu}_{3}-{\rm As}$ compound.}
\end{figure}

In Figure \ref{fig:Gruneisn}a, the Grüneisen parameter is illustrated
as a function of the parallel magnetic field $B_{\parallel}$ for
various fixed temperatures. The Grüneisen parameter shows significant
changes in response to an applied magnetic field, with the most notable
variations occurring at $B_{\parallel}\sim1$ T for temperatures around
$T\sim1$ K. As the temperature increases, the magnitude of these
changes decreases. Another region where the Grüneisen parameter becomes
relevant is around $B_{\parallel}\approx4.5$ T, exhibiting a strong
variation. However, as the temperature increases, the magnitude of
the Grüneisen parameter at this field strength decreases, eventually
diminishing. Panel (b) presents an equivalent quantity obtained by
applying a perpendicular magnetic field $B_{\perp}$, yielding results
equivalent to the previous case. Additionally, in panel (c), we depict
the variation of $\Gamma$ as a function of temperature for different
external parallel magnetic fields. A significant change in the Grüneisen
parameter $\Gamma_{\parallel}$ is observed for temperatures below
$T\sim2$ K. When the magnetic field is lower than $B_{\parallel}\sim2$
T, $\Gamma_{\parallel}$ is positive, whereas for $B_{\parallel}\gtrsim2$
T, this parameter becomes negative. Moreover, for temperatures $T\gtrsim2$
K, the Grüneisen parameter decreases significantly. The final panel
is similar to panel (c), but for a perpendicular magnetic field. Some
differences arise, such as a stronger $\Gamma_{\perp}$ compared to
the parallel case for low magnetic fields ($B_{\perp}\sim0.5$ T)
and the low-temperature region. Conversely, for large magnetic fields
($B_{\perp}\sim4.0$ T), the Grüneisen parameter ($\Gamma_{\perp}$)
is weaker than in the parallel case. In conclusion, the study of ${\rm Cu}_{3}-{\rm As}$
reveals a significant Grüneisen parameter at around $B\sim5$ T, indicating
a prominent MCE. This finding holds crucial implications for the selection
and design of magnetic refrigeration systems.

\begin{figure}
\includegraphics[scale=0.6]{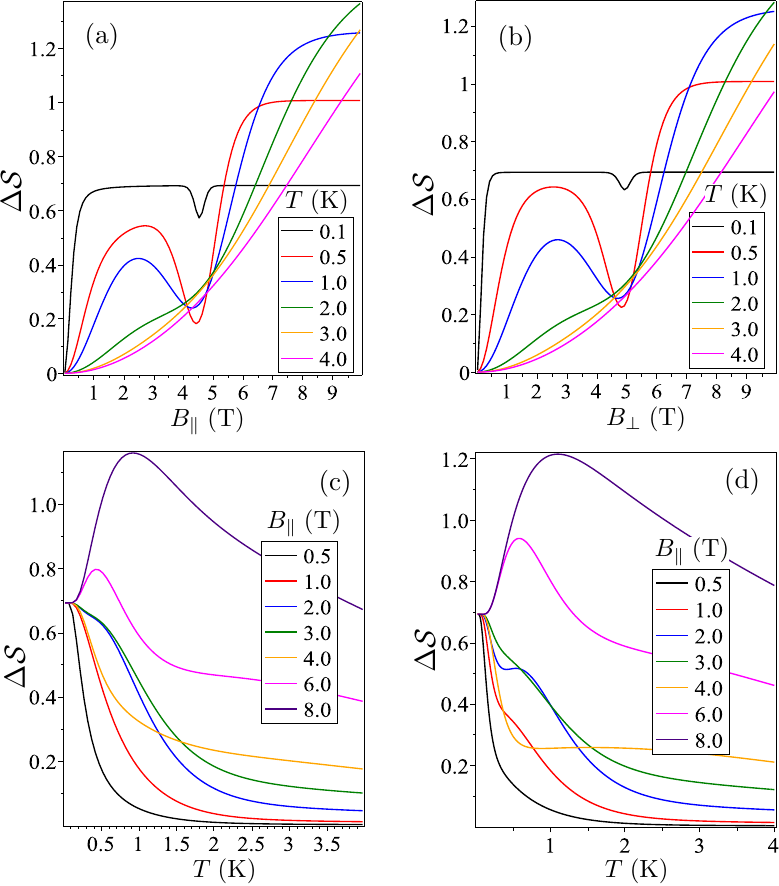}\caption{\label{fig:Delta-S}(a) Entropy variation $\Delta\mathcal{S}$ as
a function of parallel magnetic field $B_{\parallel}$, for a set
of fixed temperatures. (b) Entropy variation $\Delta\mathcal{S}$
as a function of perpendicular magnetic field. (c) $\Delta\mathcal{S}$
as a function of temperature for a variety of parallel magnetic field.
(d)$\Delta\mathcal{S}$ as a function of temperature for different
perpendicular magnetic fields.}
\end{figure}

Last but not least, the MCE can also be analyzed through the variation
in entropy, $\Delta\mathcal{S}=\mathcal{S}(0,T)-\mathcal{S}(B,T)$,
associated with the magnetic phase. This variation occurs due to the
alignment or realignment of spins, resulting in changes in the disorder
and order of the ${\rm Cu}_{3}-X$ compound and leading to a temperature
change. Therefore, $\Delta\mathcal{S}$ is an important quantity to
explore the magnetocaloric performance.

In Fig.\ref{fig:Delta-S}a, we illustrate $\Delta\mathcal{S}$ as
a function of the parallel external magnetic field $B_{\parallel}$
for several fixed temperatures. We observe that at a temperature of
$T\sim0.1$ K, the entropy remains almost constant ($\Delta\mathcal{S}\sim\ln(2)\approx0.7$).
This is due to the system being roughly double degenerate at null
magnetic field, and the degeneracy is broken by the presence of a
magnetic field. There is a slight depression at $T\approx4.5$ K,
indicating a change in the dominant phases at this magnetic field.
This behavior changes significantly as the temperature increases.
For $B_{\parallel}\lesssim4.5$ T, $\Delta\mathcal{S}$ decreases
significantly, while for $B_{\parallel}\gtrsim4.5$ T, it becomes
larger. Panel (b) shows an analogous behavior but for the perpendicular
external magnetic field $B_{\perp}$. The only difference is that
the depression occurs at $B_{\perp}\lesssim5.0$T. Furthermore, in
panel (c), we present $\Delta\mathcal{S}$ as a function of temperature
for several fixed parallel magnetic fields $B_{\parallel}$. For magnetic
fields below $B_{\parallel}\lesssim4.5$ T, $\Delta\mathcal{S}$ decreases
monotonically. However, for stronger magnetic fields $B_{\parallel}\gtrsim4.5$
T, a maximum appears, indicating a peak in $\Delta\mathcal{S}$. Similarly,
panel (d) depicts $\Delta\mathcal{S}$ as a function of temperature,
assuming a fixed perpendicular magnetic field. The behavior is mainly
similar to panel (c), although $\Delta\mathcal{S}$ does not decrease
monotonically. Additionally, for strong magnetic fields, it also exhibits
a maximum, as observed in the previous panel.

\section{Conclusions}

In this paper, we conduct a theoretical exploration of the ${\rm Cu}_{3}-X$
antiferromagnetic spin system (where $\mathrm{X=As,Sb}$), which is
identified by its isosceles or slightly distorted equilateral triangular
configurations, as detailed in reference \citep{choi06,choi08,choi12}.
This system can be accurately depicted using the Heisenberg model
on a triangular structure, incorporating factors like the exchange
interaction, Dzyaloshinskii-Moriya interaction, g-factors, and external
magnetic fields.

Recently, ${\rm Cu}_{3}-X$ has garnered significant attention due
to its fundamental properties \citep{choi06,choi08,choi12}. Furthermore,
the scientific community has shown a growing interest in the exploration
of several magnetic compounds \citep{Bouammali,Spielberg,belinsky,Robert,boudalis,stowe,kortz}
due to their diverse potential in areas such as spintronics, nanotechnology,
and biomedicine.

Our investigation uses a numerical approach to analyze both zero-temperature
and finite-temperature behaviors of the ${\rm Cu}_{3}$-like spin
system. At zero temperature, the system exhibits twofold degenerate
energy in the absence of a magnetic field and a $1/3$ quasi-plateau
magnetization when the magnetic field is varied. At finite temperatures,
our focus primarily lies on analyzing magnetic properties such as
magnetization, magnetic susceptibility, entropy, and specific heat.

In addition, we examine the MCE in relation to an externally applied
magnetic field, oriented both parallel and perpendicular to the plane
of the triangular structure. The ${\rm Cu}_{3}-X$ displays remarkably
consistent behavior for both orientations of the magnetic field. We
also extend our study to include the evaluation of the isentropic
curve, the Grüneisen parameter, and the variation in entropy during
the application or removal of the magnetic field. Therefore, in the
low temperature region below $T\sim1$K and for approximately 4.5T
and 5T for parallel and perpendicular magnetic fields, respectively,
our results confirm that the MCE is more prominent in this region.
This study could contribute to the research and development of nano-compounds
with triangular structures, potentially improving the performance
of the Magnetocaloric Effect (MCE). Such advancements may be especially
intriguing for applications in the cryogenic temperature range that
utilize moderate magnetic fields.
\begin{acknowledgments}
G. A. A. thanks CAPES, O. R. and S. M. de Souza thanks CNPq and FAPEMIG
for partial financial support. A. S. M also thanks FAPEMIG (APQ-01294-21)
for the partial funding.
\end{acknowledgments}

\end{document}